\documentclass{aa} 

\usepackage{graphicx}
\usepackage[varg]{txfonts}
\usepackage[percent]{overpic}
\usepackage{color}
\usepackage{multirow}
\usepackage{natbib,twoopt} 
\usepackage[colorlinks=true,urlcolor=blue,linkcolor=red]{hyperref} 
\bibpunct{(}{)}{;}{a}{}{,} 
\begin{document}
   
   \title{Convective blueshifts in the solar atmosphere}
   \subtitle{\textrm{II.} High-accuracy observations of the \ion{Fe}{I}\,6173.3\,\AA\ line and deviations of full-disk Dopplergrams}
   
   \author{F. Stief\inst{\ref{inst_kis}} \and J. L\"ohner-B\"ottcher\inst{\ref{inst_kis}} \and W. Schmidt\inst{\ref{inst_kis}} \and T. Steinmetz\inst{\ref{inst_mpq},\ref{inst_menlo}} \and R. Holzwarth\inst{\ref{inst_mpq},\ref{inst_menlo}}}
    \institute{
        Kiepenheuer-Institut f\"ur Sonnenphysik, Sch\"oneckstr. 6, 79104 Freiburg, Germany\\ \email{jlb@leibniz-kis.de}\label{inst_kis} \and
        Max-Planck-Institut f\"ur Quantenoptik, Hans-Kopfermann-Strasse 1, 85748 Garching, Germany\label{inst_mpq} \and
        Menlo Systems GmbH, Am Klopferspitz 19, 82152 Martinsried, Germany\label{inst_menlo}}
   \date{Received 30 Oct 2018 / Accepted xx xxx xxxx}

  \abstract 
  {Granular convective motions reach into the lower solar atmosphere, typically causing photospheric spectral lines to exhibit a differential line shift.
This Doppler shift to shorter wavelength is commonly known as convective blueshift.}
  {Spectroscopic high-accuracy measurements shall provide us with a refined determination of the absolute convective blueshift and its atmospheric distribution from disk center to the solar limb.}
  {We performed systematic observations of the quiet Sun with the Laser Absolute Reference Spectrograph (LARS) at the German Vacuum Tower Telescope. The solar disk was scanned along the meridian and the equator, from the disk center toward the limb. The solar spectrum around 6173\,\AA\ was calibrated with a laser frequency comb on an absolute wavelength scale with an accuracy of a few m\,s$^{-1}$. We applied a bisector analysis on the spectral lines to reveal the changes of convective blueshift and line asymmetry at different heliocentric positions.}
  {Being a signature for convective motions, the bisector curve of \ion{Fe}{I}\,6173.3\,\AA\ describes a ``C''-shape at disk center. When approaching the solar limb, the bisector transforms into a ``\textbackslash''-shape. The analysis of the time- and bisector-averaged line shifts yields three distinct results. Firstly, the center-to-limb variation of Doppler velocities measured with LARS reveals a significant discrepancy (up to $\mathrm{200\,m\,s^{-1}}$) to the full-disk Dopplergrams of the Helioseismic and Magnetic Imager (HMI). Secondly, we obtained a significant decrease of convective blueshift toward the solar limb. Thirdly, the line-of-sight effect of solar activity, including $p$-mode oscillations and supergranular flows, leads to a scatter of up to $\mathrm{\pm100\,m\,s^{-1}}$ at intermediate heliocentric positions.}
  {The accurate observation of the absolute convective blueshift with LARS allows the identification of systematic discrepancy with Doppler velocities measured by HMI. The center-to-limb variation of HMI suffers from an additional blueshift for $\mu<0.9$ that is incompatible with our results. LARS measurements can be taken as reference for the correction of systematic errors in the synoptic HMI Dopplergrams.}

  \maketitle
  \titlerunning{Convective blueshifts in the solar atmosphere. II} 
  \authorrunning{Stief et al.}

\section{Introduction}\label{sec_intro}
The well-known pattern of solar granulation is caused by convective overshoot from the convection zone reaching into the visible photosphere. The upflowing hot gas within the bright granular cell rises up into the photosphere in which it cools down with increasing atmospheric height. The plasma distributes horizontally toward the intergranular lanes and finally sinks back into the convection zone. Observing the spatially resolved granulation at the disk center, the upflowing granular plasma shifts spectral lines to shorter wavelength. This negative Doppler shift is called a blueshift, by convention. Accordingly, the downflow in the intergranular lanes causes a redshift toward longer wavelength. 

The spatially unresolved (or averaged) observation of a photospheric quiet Sun region typically results in an overall blueshift, commonly called the convective blueshift. {Hot rising granules induce blueshifted line profiles which, due to their brightness, contribute a greater statistical weight to the average profile, than the fainter contributions of redshifted profiles from the cooler intergranular lanes. But typically the convective blueshift is not constant along the line profile. The superposition of many profiles from various photospheric inhomogeneities produce a characteristic line asymmetry \citep{1982ARA&A..20...61D,2005ARA&A..43..481A}. In most cases, the bisector of the line profile resembles a distinct ``C''-shaped curve. Evidently, the intergranular redshifts make a stronger contribution to the outer line wings, while the granular blueshifts have a stronger impact on the inner line wings. However, there is no distinct allocation of a specific part of an averaged profile or bisector to a particular vertical height in the atmosphere. The bisector rather reflects the statistical distribution function of lateral atmospheric inhomogeneities than local vertical velocity gradients. A more detailed interpretation of convective motions in terms of rough atmospheric levels would require complex three-dimensional simulations.}

The amount of convective blueshift also depends significantly on the observed position on the solar disk. At the center of the solar disk, the convective blueshift reaches some hundred $\mathrm{m\,s^{-1}}$. Toward the solar limb, the velocities decrease significantly and can even turn into a redshift \citep{1984SoPh...93..219B,1985SoPh...99...31B}. {This center-to-limb variation originates from combinations of line-of-sight opacity effects (sampling higher atmospheric levels near the limb), and the changing projection angle of granular motions (vertical motions contributing less Doppler shifts near the limb).}

Another aspect of the center-to-limb variation of the convective blueshift is suggested to arise from the horizontal flow fields of the solar granulation \citep{1984SoPh...93..219B,1985SoPh...99...31B}. Toward heliocentric positions of around $\mu=0.75$, the convective blueshift can increase by a few ten $\mathrm{m\,s^{-1}}$ due to the more effective sampling of the horizontal velocity component of granular flows. This trend was also confirmed by numerical syntheses \citep{2011A&A...528A.113D}. 
 
This work is the second part of the publication series ``Convective blueshifts in the solar atmosphere''. The series concentrates on the determination of absolute convective blueshifts of the most frequently used spectral lines in the visible range of the solar spectrum. The observations were performed with the Laser Absolute Reference Spectrograph \citep[LARS,][]{Doerr2015,2017A&A...607A..12L}. In order to obtain absolute wavelengths and Doppler velocities at the $\mathrm{m\,s^{-1}}$ accuracy, the solar spectra are calibrated with a laser frequency comb. By providing precise and comprehensive spectroscopic measurements of the quiet Sun convective blueshift, including the line asymmetry and its center-to-limb variation, we aim to reach a better understanding and three-dimensional image of the convective transmission into the solar atmosphere. 

In the first article \citep[][hereafter referred to as Paper I]{2018A&A...611A...4L} of the series, we refined the amount of convective blueshift of the spectral lines around 6302\,\AA\ and investigated their line asymmetry, as well as their center-to-limb variation. In this work, we follow the same approach as in Paper I, but apply the analysis on the observations taken in the 6173\,\AA\ range. An upcoming third article will combine the observed convective blueshifts of a set of important lines in the visible part of the solar spectrum.

The presented accurate reference values for line-of-sight velocities allow for comparisons with other solar spectrometers and polarimeters, and possibly the calibration of absolute velocities. In this work, we will draw a comparison between our LARS observations of \ion{Fe}{I}\,6173.3\,\AA\ and the Dopplergrams from the Helioseismic and Magnetic Imager \citep[HMI,][]{2012SoPh..275..229S}, on board of the Solar Dynamics Observatory (SDO), obtained with the same spectral line. So far, HMI can only provide relative velocities which show temporal and spatial large-scale variations of the order of several hundred $\mathrm{m\,s^{-1}}$ \citep{Loehner+Schlichenmaier2013}. Though, an accurate calibration of Doppler velocities is essential for the determination of small-scale up- and downflows or sunspot flows. We argue that a recalibration of HMI Dopplergrams with LARS reference values provides an accuracy of better than $\mathrm{100\,m\,s^{-1}}$ for absolute HMI velocities all across the solar disk.

\section{Observations}\label{sec_data}
The observations were performed in 2016 during August and October with the LARS instrument at the German Vacuum Tower Telescope (VTT) on Tenerife.

\subsection{LARS observations}\label{susec_LARS_meas}
The measurements and the data calibration was done in the same manner as presented in Paper I \citep{2018A&A...611A...4L}. In this section, we give a short overview on the most important observational characteristics, and refer the reader to Paper I for a more detailed description. Distinctions to the observations of Paper I are explicitly stated hereafter.

\begin{figure}[htpb]
\includegraphics[width=\columnwidth]{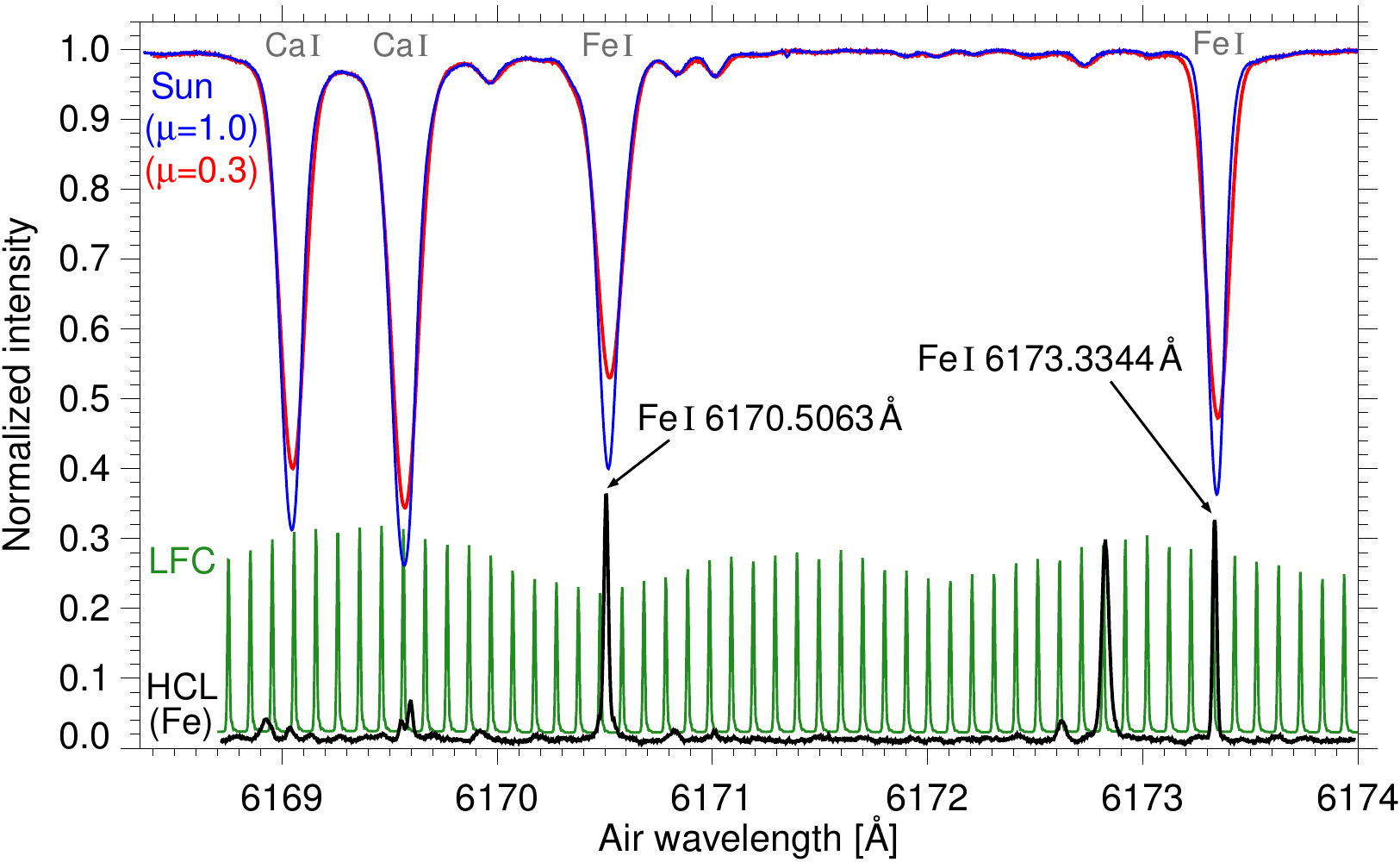}
\caption{Observed spectra in the wavelength region around 6173\,\AA. The quiet Sun absorption spectra were observed at the disk center ($\mu=1.0$, blue curve) and close to the solar limb ($\mu=0.3$, red curve). The ions are stated in gray color. The emission spectra of the laser frequency comb (LFC, green curve) and the iron hollow cathode lamp (HCL, black curve) were measured to calibrate absolute wavelengths and Doppler shifts. The \ion{Fe}{I} reference lines are marked.}
\label{fig_spectral_profiles}
\end{figure}

\begin{figure}[htpb]
\includegraphics[width=\columnwidth]{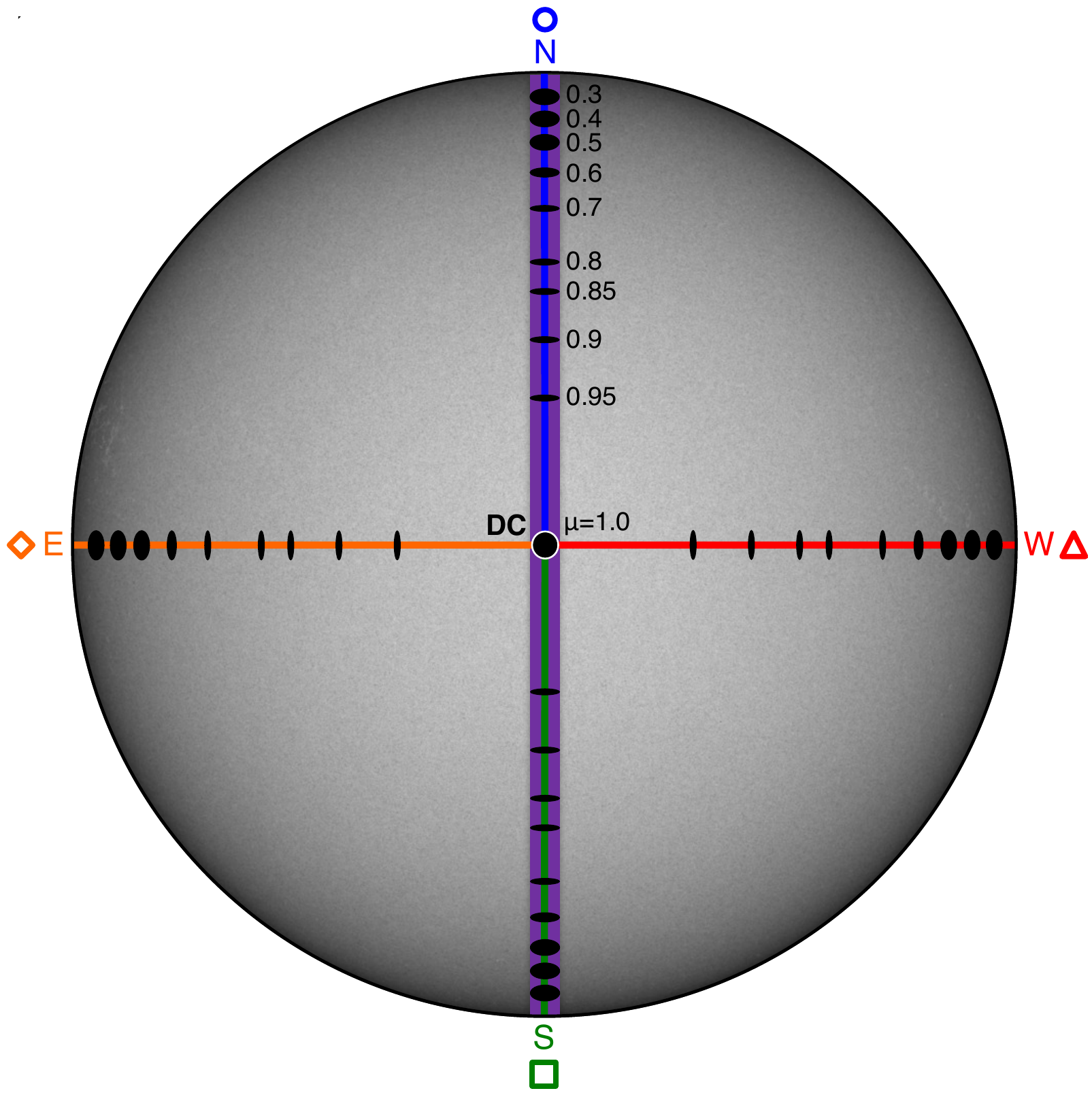}
\caption{Observation scheme. We scanned ten heliocentric positions ($\mu=\cos\theta$) from the disk center (DC, $\mu=1.0$) toward the solar limb ($\mu=0.3$) along the meridian (N-S) and equator (E-W): the north (blue, circle), south (green, rectangle), east (orange, diamond), and west (red, triangle) radian. The black ellipses indicate the field of view of LARS, the purple rectangle marks the HMI region for comparison.}
\label{fig_clv_sun}
\end{figure}

\paragraph{Measurement:}
The observations were performed with LARS \citep{Doerr2015,2017A&A...607A..12L}. The instrument is a combination of VTT's high-resolution echelle spectrograph ($\lambda/\Delta\lambda > 700\,000$, at $\lambda\sim6173\,\AA$) for spectroscopic measurements and a laser frequency comb for an absolute wavelength calibration. The solar absorption spectrum of the 6173\,\AA\ range and the emission spectrum of the frequency comb (both shown in Fig.\,\ref{fig_spectral_profiles} were recorded in an alternating cycle. The exposure time for each spectrum was 0.5\,s, the total cycle time for both spectra was set to 1.5\,s. To account for solar $p$-mode activity, each observation sequences contained 800 consecutive cycles, making a time sequence of 20\,min. We selected the 10\arcsec\, field aperture to integrate the solar light. To study the center-to-limb variation of the \ion{Fe}{I}\,6173.3\,\AA line, we performed a systematical scan of the solar disk. The applied scanning scheme is displayed in Fig.\,\ref{fig_clv_sun} (adapted from Fig.\,3 of Paper I). We chose ten heliocentric positions $\mu=\cos \theta$, from the disk center ($\mu=1.0$) to close proximity to the solar limb ($\mu=0.3$). At each of the 40 solar disk positions, at least one sequence was observed. Using LARS G-band context images and HMI full-disk magnetograms, we verified that only quiet Sun regions were scanned. As described in Paper I, we additionally applied an oscillatory motion of the telescope pointing around the nominal position. The circular (or elliptical) oscillation was performed with a frequency of 5\,Hz and an amplitude of up to 15\arcsec. This method ensured a more effective averaging over acoustic oscillations and especially supergranular horizontal flows.

\paragraph{Calibration:} The data calibration was performed with the pipeline developed by \citet{Doerr2015}. A schematic overview was given by \citet{2017A&A...607A..12L}. The optimized calibration included a number of steps to reach an accuracy of the order of a few $\mathrm{m\,s^{-1}}$. First, dark and flat-field correction was applied to all spectra. Then, each solar spectrum was calibrated with the proximate emission spectra of the laser frequency comb. The resulting instrumental accuracy reached a level of a few femtometer, which translates into a Doppler velocity accuracy of $\mathrm{1\,m\,s^{-1}}$. In the next step, we reduced all orbital, radial and rotational motions of the Sun and Earth, with a given uncertainty of the model of around $\mathrm{0.1\,m\,s^{-1}}$. We used the ephemerides code developed by \citet{Doerr2015}, based on NASA's SPICE toolkit \citep{Acton1996}. We used the spectroscopic differential rotation model by \citet{1990ApJ...351..309S}. Given by a telescope pointing accuracy of 1--2\arcsec, we estimate total velocity error from the solar rotation for the nominal heliographic positions to $\mathrm{4\,m\,s^{-1}}$. A summarized error discussion is given at the beginning of Section\,\ref{sec_results}. Finally, the spectral lines should only be affected by the {constant gravitational redshift} and the local solar activity itself. The latter comprises convective motions and acoustic oscillations of the Sun. By temporal averaging of the 20\,min sequences, we can diminish the scatter, caused by oscillations, of the average Doppler shift to a few $\mathrm{m\,s^{-1}}$. From the propagation of errors, we obtain a total uncertainty of around $\mathrm{5\,m\,s^{-1}}$ including all stated systematic and statistical errors.

We focused our investigations on the neutral iron line at 6173.3\,\AA. In the present work, we are only interested in line shifts whereas the high magnetic sensitivity is not relevant here. To minimize the influence of magnetic fields on the line profiles, we avoided regions with apparent magnetic activity. The line profile is not deformed by atomic or molecular blends, nor by telluric lines. The formation height covers the lower photosphere, up to atmospheric heights of around 280\,km \citep{1991sopo.work.....N} above the optical depth unity at 5000\,\AA. The \ion{Fe}{I}\,6173.3\,\AA\ is well suited for measurements of Doppler velocities in the quiet Sun \citep{2005A&A...439..687C}, and for bisector {analyses.}

The calculation of Doppler shifts ($\Delta\lambda=\lambda-\lambda_0$) requires an accurate wavelength reference. To provide the best possible accuracy, we measured the laboratory wavelength $\lambda_0$ of \ion{Fe}{I}\,6173.3\,\AA\ with the iron hollow cathode lamp of LARS. The emission spectrum of the lamp is shown in Fig.\,\ref{fig_spectral_profiles} with an adapted intensity scale for better illustration. The central wavelength was fitted with a Voigt function to the respective emission line. We yield a laboratory wavelength of $\mathrm{\lambda_0=6173.3344\,\AA}$, with an uncertainty of less than $\mathrm{0.1\,m\AA}$ (translated into a velocity error of only a few $\mathrm{m\,s^{-1}}$). In comparison, the observed air wavelength by \citet{1994ApJS...94..221N}, provided by the National Institute of Standards and Technology Atomic Spectra Database \citep[NIST ASD,][]{NIST_ASD}, amounts to 6173.3352\,\AA, {with an uncertainty of $\mathrm{1.1\,m\AA}$.}

We note that throughout the paper we adopted common practice and used air wavelengths when referring to observed wavelengths.

\subsection{Bisector analysis}\label{susec_bisectors}
\begin{figure}[htpb]
\includegraphics[trim= 5mm 0mm 3mm 0mm, clip, width=\columnwidth]{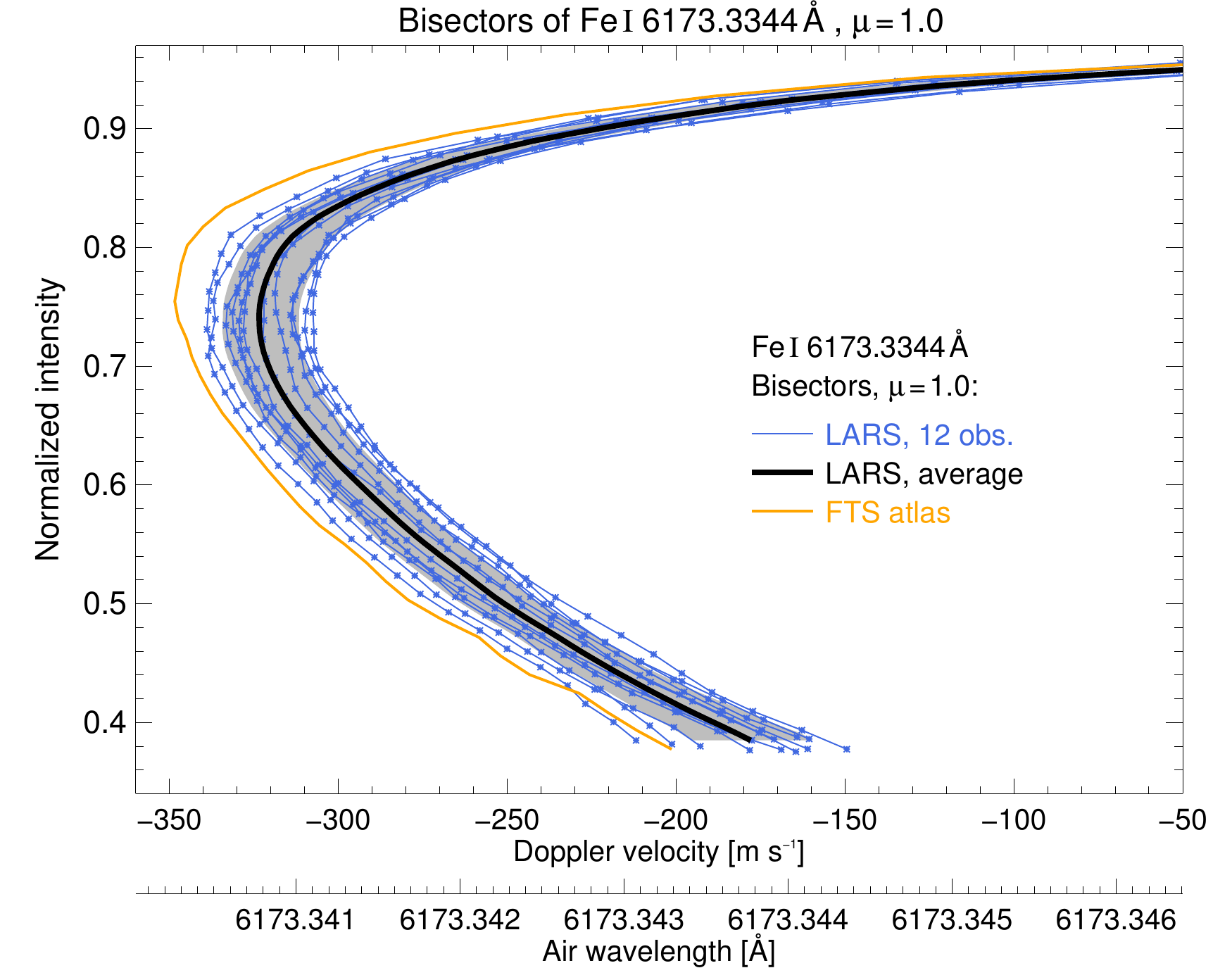}
    \caption{Bisectors of \ion{Fe}{I}\,6173.3\,\AA\ measured at the solar disk center ($\mu=1.0$). The normalized intensity is displayed against the absolute air wavelength (lower x-axis) and the translated Doppler velocity (upper x-axis). The individual bisectors of 12 observational sequences are shown as blue curves with asterisks. The average bisector (black solid line) and its standard deviation (gray area) are given. The {line} bisector from the Fourier Transform Spectrometer \citep[FTS,][]{1999SoPh..184..421N} {disk-center} atlas is added in orange.}
    \label{fig_bisectors_6173_mu10}
\end{figure}

To study the exact position and shape of the spectral line, we performed a bisector analysis. The center of horizontal line segments at various depths along the absorption profile yields the detailed asymmetry of the line shape, and thus provides us with the necessary height-dependent gradient of Doppler shifts. We calculated bisectors at intensities ranging from the line minimum to an upper threshold level of 95\% of the continuum intensity. We used an equidistant step size of 3\,\% of the normalized line depth. We thus avoided oversampling and inclusion of statistical fluctuations. The obtained bisector curves of \ion{Fe}{I}\,6173.3\,\AA\ at solar disk center are shown in Fig.\,\ref{fig_bisectors_6173_mu10}.

The observed air wavelengths $\lambda$ were transformed into Doppler velocities $\mathrm{v_{los}}$ by
\begin{equation}
\mathrm{v_{los}=c\cdot(\lambda-\lambda_0)/\lambda_0-v_{grs}}\label{eq1}
\end{equation}
with the speed of light $\mathrm{c}$, the reference rest wavelength $\mathrm{\lambda_0}$, and the {total gravitational redshift $\mathrm{v_{grs}}$. For ground-based observations, the latter sums up to $\mathrm{+634.85\,m\,s^{-1}}$. In the gravitational field of the Earth, the redshift from the solar photosphere (to infinity) of $\mathrm{+636.31\,m\,s^{-1}}$ reduces to $\mathrm{+633.35\,m\,s^{-1}}$. Neglecting tiny variation of the orbital speed of the Earth ($\mathrm{30\,km\,s^{-1}}$) around the Sun, the transverse Doppler effect induces an additional relativistic redshift of $\mathrm{+1.5\,m\,s^{-1}}$.}

\subsection{HMI observations}\label{susec_HMI_meas}
We used HMI full-disk Dopplergrams that were summed over 12 minutes. The radial velocity of the SDO satellite has been removed \citep[code provided by M. Schmassmann on the basis of][]{2016ApJ...823..101S}. To perform a comparison with our LARS observation, we computed the average curve of the convective blueshift along the meridian (north-south). At each heliocentric angle, we averaged the values within an 61\,pixel wide region in equatorial direction (east-west). To obtain the radial slope, we averaged the north and south axis of the meridian. 

\section{Results and discussion}\label{sec_results}
We calculated the bisector profiles at each instant in time. The average shift of the bisector profile was defined as the Doppler shift of the spectral line. With a temporal cadence of 1.5\,s and time sequences of 20\,min, we were able to capture the basic solar p-mode oscillations and fast atmospheric variations. In accordance with Paper I, we find the same evolution of decreasing oscillations from the disk center toward the solar limb. At the disk center, the dominating 5\,min-oscillation is most prominent with amplitudes of up a few hundred $\mathrm{m\,s^{-1}}$. In comparison to other studies \citep[e.g. references given in ][]{2002tsai.book.....S}, the measured amplitude is smaller since we averaged spatially over an area of 30\,\arcsec. With increasing distance to the disk center, we yield decreasing oscillatory amplitudes. At heliocentric positions close to the solar limb ($\mu<0.5$), the temporal oscillation has almost vanished. This is largely caused by the increasing angle of incidence between the line-of-sight and the dominating vertical component of the oscillation. But additionally, the observed atmospheric height increases for large heliocentric angles. By the more tangential line-of-sight, the spectral line samples higher atmospheric layers.

We removed the oscillatory component by computing the temporal average for each 20\,min sequence. The time-averaged bisector profile and its mean shift provided a good estimator for the Doppler velocity caused by the granular convection. The systematic error of the average Doppler velocity is composed of instrumental errors ($\mathrm{1\,m\,s^{-1}}$), the uncertainty of the ephemeris correction ($\mathrm{0.1\,m\,s^{-1}}$), the telescope pointing accuracy of 1-2\arcsec\ and the uncertainty of the applied solar rotation model ($\mathrm{4\,m\,s^{-1}}$), and the uncertainty of the reference wavelength (around $\mathrm{3\,m\,s^{-1}}$ for the \ion{Fe}{I} lines). Statistical fluctuations of the average line shift did arise from solar $p$-mode oscillations and, to a minor fraction, statistical readout scatter. The observation of sequences with several hundred time steps reduced the error of the mean convective blueshift. The standard deviation of a single measurement has a typical amount of $\mathrm{50-70\,m\,s^{-1}}$. Dividing by the square root of the number of cycles (up to 800) decreases the error of the average Doppler shift to only few $\mathrm{m\,s^{-1}}$. By means of error propagation, we obtain an overall error of around $\mathrm{10\,m\,s^{-1}}$ for the average Doppler shift of each 20\, min observation.

In the following sections, the presented analysis was performed with the temporally averaged bisectors at each heliocentric position. In Section\,\ref{susec_LARSHMI}, we will compare the center-to-limb variation of the convective blueshift measured with LARS and HMI. A more detailed analysis of the convective blueshift observed with LARS is given in Section\,\ref{susec_clv}.

\subsection{Comparison of center-to-limb variation between LARS and HMI}\label{susec_LARSHMI}

The Helioseismic and Magnetic Imager performs continuous spectro-polarimetric observations using the \ion{Fe}{I}\,6173.3\,\AA\ line. This provides us with the opportunity of drawing a comparison between HMI and LARS measurements. However, we must note the different instrumental specifications which make a one-to-one comparison to be treated with caution. To analyze the spectral line, HMI performs a spectrometric scanning at six equidistant wavelength positions in the range of $\mathrm{\pm172.5\,m\AA}$, symmetrically distributed around the target wavelength of the line center at 6173.3433\,\AA\ \citep{2012SoPh..275..229S}. With a full-width at half-maximum of $\mathrm{76\,m\AA}$ of the instrumental point-spread function, the spectral resolution amounts to around 81\,000. The line center is calculated from the discrete estimate of the first and second order Fourier coefficients \citep{2012SoPh..278..217C}. The relative Doppler shift is defined as the wavelength shift with respect to the given target wavelength.

\begin{figure}[htpb]
\includegraphics[trim= 3mm 1mm 4mm 0mm, clip, width=\columnwidth]{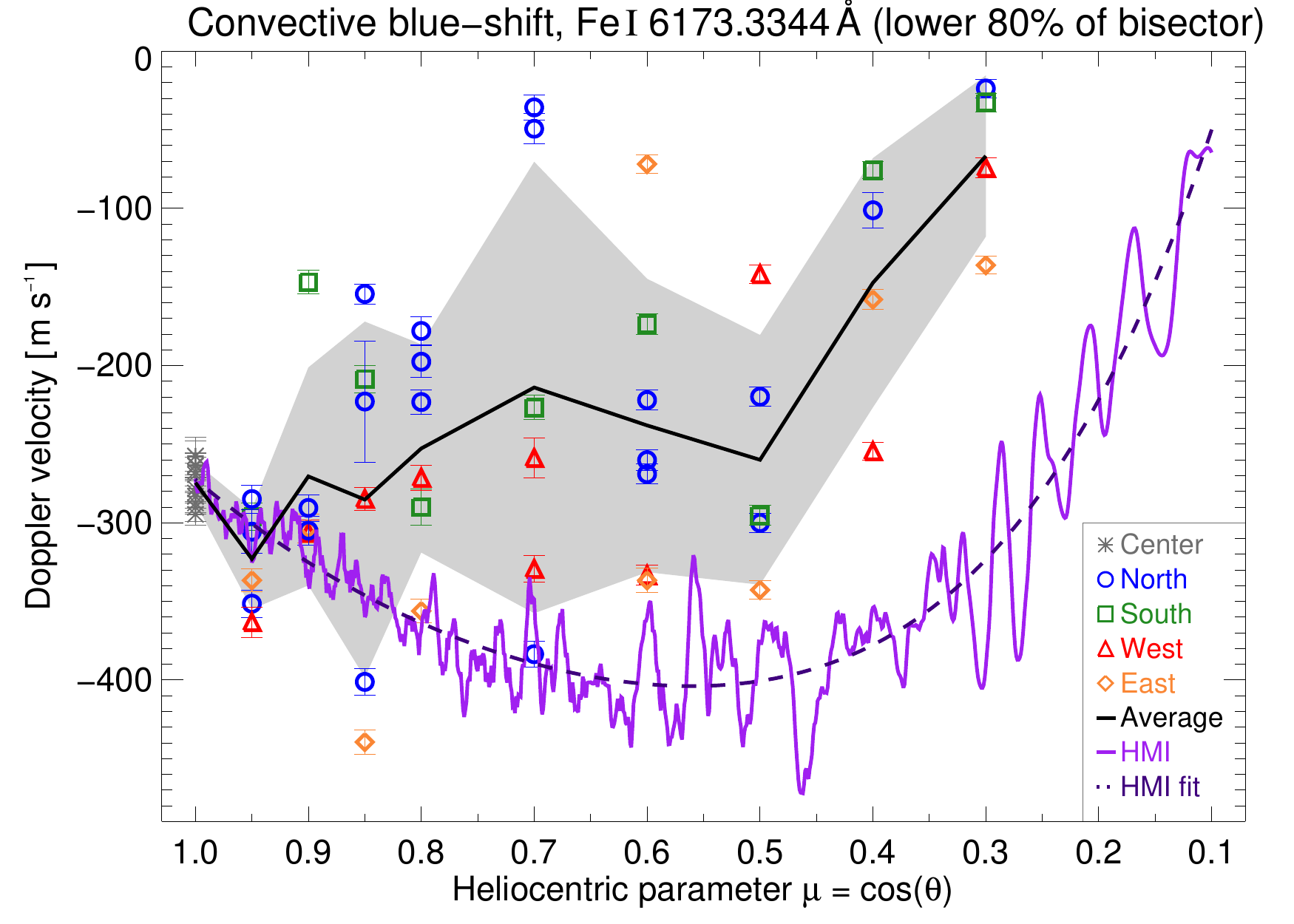}
\caption{Comparison of the center-to-limb variation of the convective blueshift of \ion{Fe}{I}\,6173.3\,\AA\ between LARS and HMI. Doppler velocities of LARS represent the average of the lower 80\,\% of the bisector profile (in contrast to the lower 5\,\% in Fig.\,\ref{COG Plot}). Each data point represents the temporal average of an observation sequence, with error bars indicating the mean error. Colors and symbols indicate the radial axis of the solar disk. The mean curve and its standard deviation are displayed as black solid line and gray area. The HMI curve (purple solid line) and its polynomial fit (purple dashed line, Eq.\,\ref{eq2}) range from the solar disk center ($\mu=1.0$) toward the limb ($\mu=0.1$).}
\label{img: HMI comparison}
\end{figure}

A comparison of LARS and HMI data involves the assimilation of their spectroscopic and observational conditions. To guarantee the validity of the spatial comparison, we selected a region along the northern and southern radial axis in the HMI Dopplergrams which includes all positions observed with LARS. We reduced the impact of supergranulation by calculating the average HMI Dopplergram for the full time period of both observing campaigns. Changes in the apparent solar radius, the position angle $\mathrm{P_0}$, and the heliographic latitude $\mathrm{B_0}$ were taken into account. A first direct comparison of the measured convective blueshift and its center-to-limb variation for LARS and HMI is shown in Fig.\,\ref{img: HMI comparison}. The Doppler velocities of LARS were calculated as the average of the lower 80\,\% of the line bisector. Since the original LARS spectrum has a resolution of more than 700\,000, whereas HMI samples the shift of the entire line profile with a much smaller resolution, we find this bisectorial average to be an adequate first approach for a comparison. Since HMI provides relative Doppler velocities, we computed the offset of both mean center-to-limb curves for the range between $\mu=1.0$ and $\mu=0.9$, and adapted the HMI curve to the absolute LARS observations. 

As displayed in Fig.\,\ref{img: HMI comparison}, the slopes of the center-to-limb curves of LARS and HMI are in good agreement only around the disk center ($\mu=1.0-0.95$). But evidently, both curves start to deviate from $\mu=0.9$ to $\mu=0.7$. The convective blueshift observed with LARS is rather stable around $\mathrm{-250\,m\,s^{-1}}$ in the range between $\mu=0.9$ and $\mu=0.5$. In the case of HMI, the convective blueshift increases monotonically by around $\mathrm{100\,m\,s^{-1}}$ within the same $\mu$-range. From $\mu=0.5$ toward the solar limb ($\mu\rightarrow0$), both center-to-limb curves have a similar gradient again, but with an offset of around $\mathrm{200\,m\,s^{-1}}$. One can argue that a relative shift of the HMI curve by $\mathrm{+200\,m\,s^{-1}}$ would reproduce the actual slope of the center-to-limb variation of LARS at $\mu\le0.7$. However, applying this offset would result in Doppler shift of HMI which are too redshifted at $\mu>0.7$.

Besides, the effect of supergranulation is starting to superimpose the general trend of the curve. Close to the disk center, the horizontal flow fields of supergranulation are perpendicular to the line of sight and, thus, do not affect the center-to-limb curve. With increasing heliocentric angle, the imprint of supergranular flows becomes apparent as an superimposing oscillatory feature. Long-term averaging over several weeks or month would reduce this scatter. In our case, we performed a 5th degree polynomial fit 
\begin{equation}
\mathrm{v_{los}}(\mu)=223\mu^5-3352\mu^4+6963\mu^3-7347\mu^2+4134\mu-893\,\mathrm{[m\,s^{-1}]}
\label{eq2}
\end{equation}
on the HMI curve. The obtained center-to-limb variation featuring an initial slight increase and successive strong decrease in convective blueshift is in line with the theory \citep{1985SoPh...99...31B} of granular motion and the effect of horizontal flows. However, the fit yields a maximum blueshift at $\mu=0.56$. Compared to theoretical syntheses \citep[e.g.,][]{2011A&A...528A.113D} which suggest that the maximum convective blueshift is reached at positions between $\mu=0.8$ and $\mu=0.7$, the center-to-limb curve of HMI seems to be significantly displaced or strongly influenced by instrumental or calibration effects. In the following section, we will investigate the cause for the difference between LARS and HMI results. 

\paragraph{Calibrating LARS data under HMI conditions:} A plausible cause for the deviation between LARS and HMI results are differences of their spectroscopic observation conditions. Another possible reason could be the different approach of calculating the Doppler velocity. To test the impact of instrumental and calibration effects on the final outcome, we performed a theoretical evaluation of our LARS observations under the instrumental conditions of HMI \citep{2012SoPh..275..229S} and the applied Doppler velocity calibration \citep{2012SoPh..278..217C}. 

\begin{figure}[htpb]
\includegraphics[trim= 3mm 0mm 2mm 0mm, clip, width=\columnwidth]{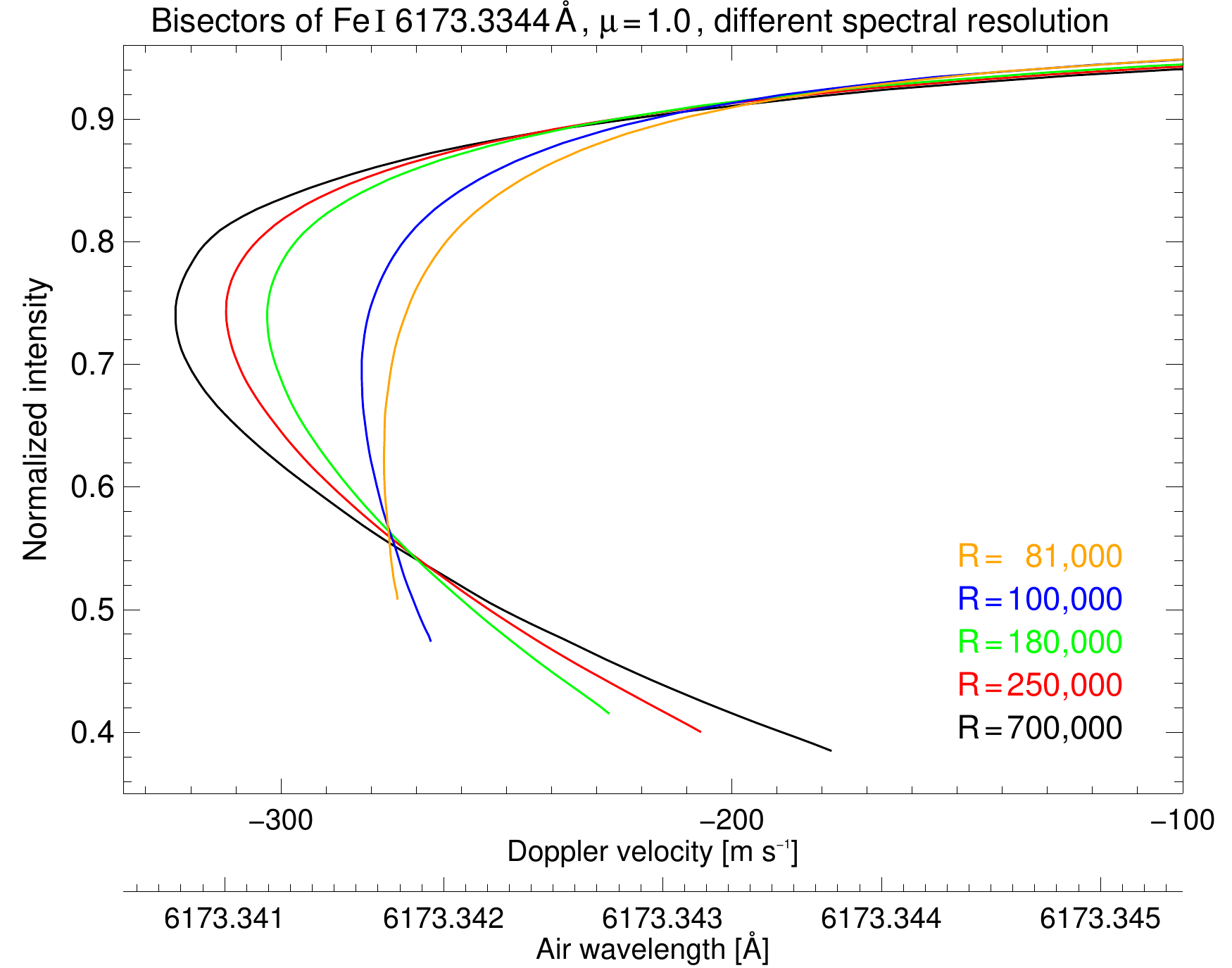}
\caption{Comparison of the bisector profiles of \ion{Fe}{I}\,6173.3\,\AA\ for different spectral resolutions. The profile measured at disk center decreases in spectral resolution from the original LARS resolution of 700\,000 (black) to the HMI resolution of 81\,000 (orange).}
\label{img: resolution comparison}
\end{figure}

\begin{figure}[htpb]
\includegraphics[trim= 0mm 1mm 0mm 0mm, clip, width=\columnwidth]{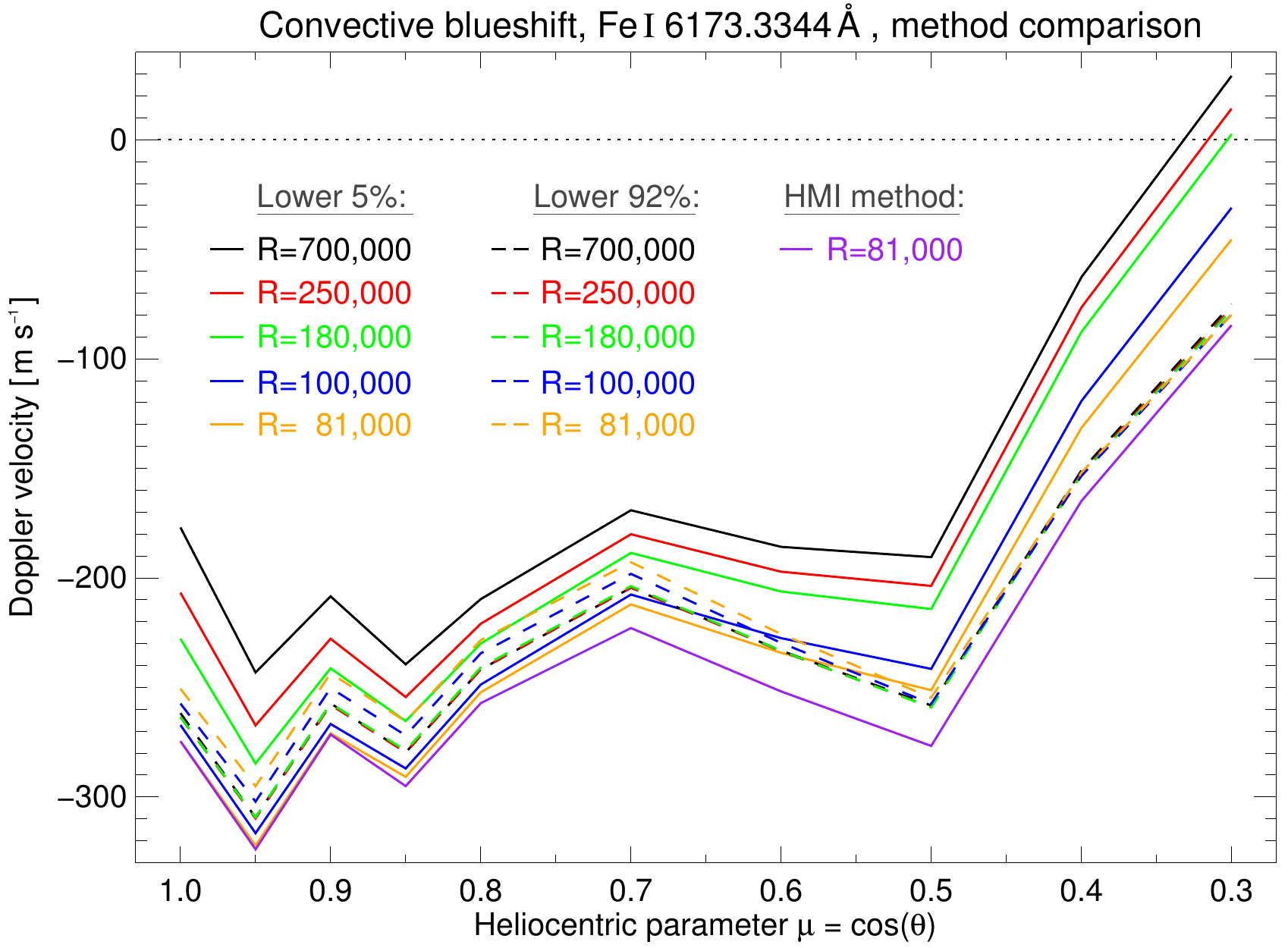}
\caption{Comparison of the convective blueshift measured with LARS from disk center toward the solar limb on the basis of different spectral resolutions $R$, line segments, and velocity calibration approaches applied to the original LARS data. The variations of the line core velocity (lower 5\,\% of the bisector) are displayed as solid lines (from black to orange). The variations under consideration of the lower 92\,\% of the line bisector are shown as dashed lines. The variation of Doppler velocities obtained with the HMI approach are drawn as purple solid line.}
\label{img: CLV LARS impact comparison}
\end{figure}

First, the resolutions of the LARS spectra was equalized to the spectral resolution of HMI. We performed a convolution of the original spectra with the Gaussian point-spread function of HMI to lower the resolution from from 700\,000 to 81\,000. To track the evolution of the line asymmetry, we included degrading steps at resolutions of 250\,000, 180\,000, and 100\,000. The theoretical evolution of the \ion{Fe}{I}\,6173.3\,\AA\ bisector at disk center is shown in Fig.\,\ref{img: resolution comparison}. With decreasing spectral resolution, the strongly C-shaped asymmetry vanishes and turns into a more uniform bisector. Moreover, the Doppler velocity of the line core increases by around $\mathrm{100\,m\,s^{-1}}$. Fig.\,\ref{img: CLV LARS impact comparison} displays the change of the center-to-limb variation of the convective blueshift for different spectral resolutions and bisector segments. The segments sampled either the line core by averaging the Doppler velocities of the lower 5\%, or the entire line by the lower 92\%\ of the bisector. In contrast to the overall increase (up to $\mathrm{100\,m\,s^{-1}}$) of the line core blueshift with decreasing spectral resolution, the center-to-limb variation of the entire line is hardly affected ($\mathrm{<15\,m\,s^{-1}}$) by the change in resolution.

As a second step, we derived the intensities of the degraded (81\,000) spectral line profile at the six nominal wavelength positions of HMI. With an equidistant step size of 68.8\,m\AA, the wavelength positions are symmetrically distributed in a range of $\mathrm{\pm172.5\,m\AA}$ around the target wavelength at 6173.3433\,\AA. Thus, none of the positions samples the minimum of the unshifted line profile. The line center is calculated according to the HMI algorithm \citep{2012SoPh..278..217C}. The discrete estimates of the first and second order Fourier coefficients were calculated by applying the intensities at the six scan positions. The average of both is taken as the line position. After subtracting the gravitational redshift, we received the absolute Doppler shift by translating the relative wavelength shift to the actual reference wavelength of 6173.3344\,\AA. Finally, wavelength shifts were transformed into Doppler velocities.

The resultant center-to-limb variation of the convective blueshift based on the HMI method is added in Fig.\,\ref{img: CLV LARS impact comparison}. The direct comparison with the original bisector analysis of LARS reveals the impact of the spectral resolution and velocity calibration approach. Evidently, the HMI approach yields the strongest convective blueshifts at all heliocentric positions. On the other extreme, the center-to-limb variation of the line core velocities of the original LARS data has always the weakest blueshift. The center-to limb variations of all other approaches lie in the range between both curves. We note that the HMI method is robust and yields results which are in surprisingly well agreement with the bisectorial analysis, considering that the approach requires only six wavelength positions of which not even one samples the line minimum. Compared to the global curves for the entire bisectorial average, the velocity difference is less than $\mathrm{30\,m\,s^{-1}}$. More astonishingly, the results of the HMI method match perfectly well with the bisector analysis of the line core velocities at a spectral resolution of 81\,000 for observations close to the solar disk center. Only from $\mu=0.8$ toward the solar limb, the deviation becomes more apparent.   

\paragraph{Comparison of the adapted LARS data with HMI:}

\begin{figure}[htpb]
\includegraphics[trim= 0mm 0mm 0mm 0mm, clip, width=\columnwidth]{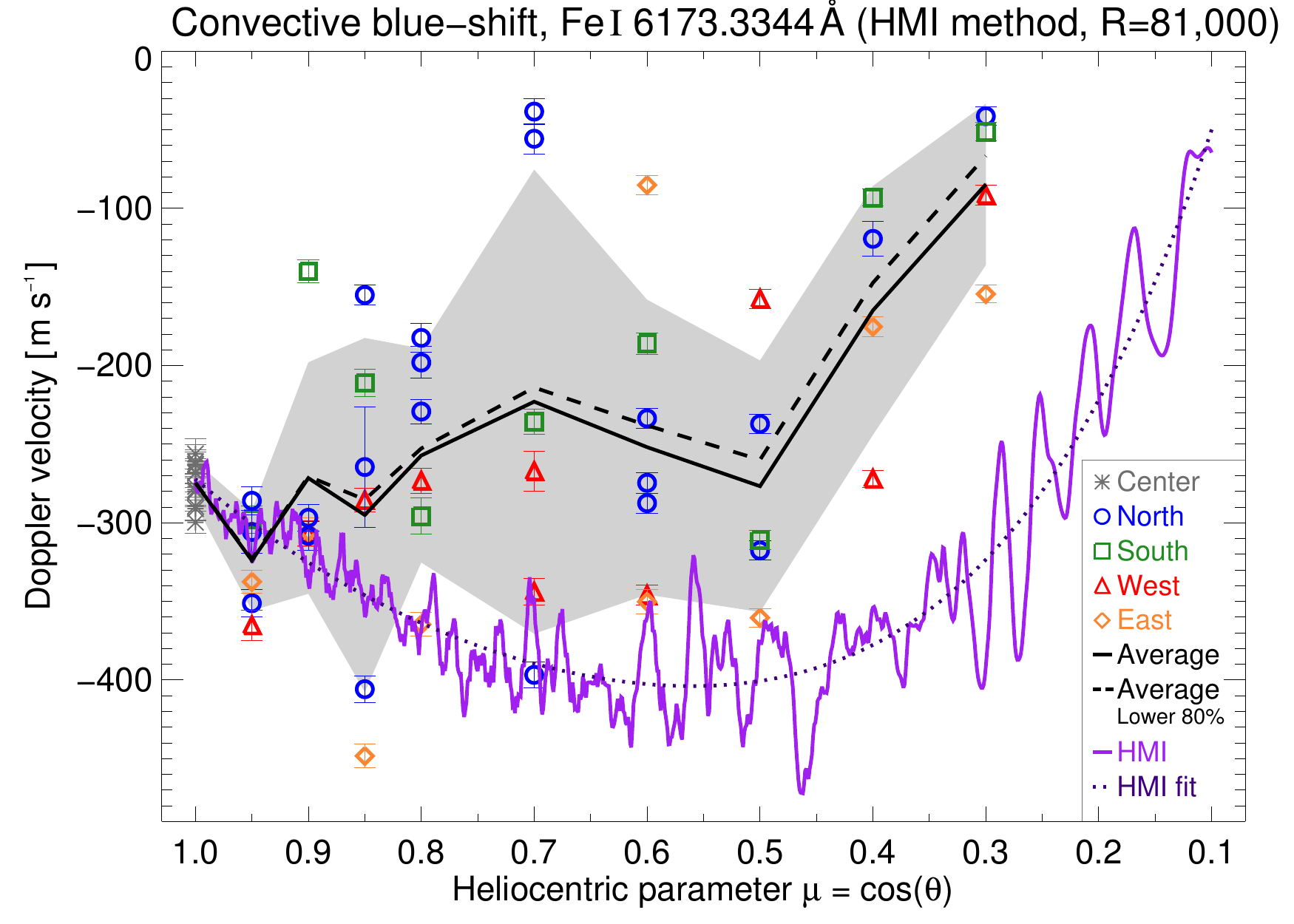}
\caption{Comparison of the convective blueshift between the adapted LARS data and HMI. Contrary to Fig.\,\ref{img: HMI comparison}, the Doppler velocities were obtained with the HMI approach on the LARS spectra at a degraded resolution of 81\,000. Colors and symbols indicate the radial axis of the solar disk. The mean curve and its standard deviation are displayed as black solid line and gray area. The original mean center-to-limb variation from Fig.\,\ref{img: HMI comparison}, obtained by the average of the lower 80\,\% of the bisector profile, is added as black dashed line. The HMI curve and its polynomial fit are drawn as purple lines.}
\label{img: HMI LARS(HMI method and average) comparison}
\end{figure}

The analysis of LARS and HMI data under the same spectroscopic properties and calibration approach allows for a direct comparison of the obtained convective blueshift and its center-to-limb variation. The re-calibrated Doppler velocities of the observation sequences are presented in Fig.\,\ref{img: HMI LARS(HMI method and average) comparison}. At the solar disk center between $\mu=1.0$ and $\mu=0.9$, the average center-to-limb variation of the adapted convective blueshift is in very good agreement with both, the original LARS curve obtained by the average of the lower 80\,\% of the bisector profile and the HMI curve. Toward the solar limb, the adapted and the original center-to-limb variation diverge slowly. But even at small heliocentric positions around $\mu=0.5$, the difference between the both average curves is less than $\mathrm{20\,m\,s^{-1}}$. We conclude that the applied velocity calibration and spectral resolution results in a slightly stronger convective blueshift of the order of a few ten $\mathrm{m\,s^{-1}}$. However, the instrumental difference and calibration approach can not explain the significant difference to the center-to-limb variation provided by HMI. With a difference of more than $\mathrm{200\,m\,s^{-1}}$ toward small heliocentric $\mu$-values, the deviation of the HMI measurements is a multiple larger than the given standard deviation of our LARS observations. Since Doppler velocities measured with LARS are accurate to the $\mathrm{m\,s^{-1}}$ scale and since we do not expect a significant change of the center-to-limb curve by increasing the statistical sample size, the deviation must arise from HMI.  

We can only speculate on intrinsic instrumental effects of HMI which could corrupt the measurement of solar Doppler velocities. The emerging wavelength dependence described by \citet{2012SoPh..275..285C} could be explained by systematic wavelength shifts across the field-of-view of the HMI interferometer, and thus, a variation across the solar disk. The variation of the light path through the interference filters of HMI causes a ring-shaped wavelength dependence. Those instrumental interference fringes, reported by \citet{Loehner+Schlichenmaier2013} and Fig.\,5 therein, can lead to systematic large-scale velocity variations of the order of $\mathrm{\pm100\,m\,s^{-1}}$ across the solar disk. We believe that a systematic reduction of those interference fringes would lead to a change of the slope of the center-to-limb variation in HMI Dopplergrams toward a reproduction of the actual center-to-limb variation of the convective blueshift measured by LARS. We are confident that increasing the number of LARS observations will reduce the scatter and smoothen the mean center-to-limb curve of the convective blueshift. At this point, a calibration of HMI Dopplergrams according to the reference velocity given by LARS would allow a transformation of the given relative Doppler velocities of HMI into absolute Doppler velocities with an accuracy of $\mathrm{\pm100\,m\,s^{-1}}$. We argue that this accuracy can be increased to a few ten $\mathrm{m\,s^{-1}}$ by the reduction of the intrinsic systematic wavelength errors of HMI. 

\subsection{High-resolution observations of the \ion{Fe}{I}\,6173.3\,\AA\ convective blueshift}\label{susec_clv}
In this section, we want to highlight the analysis of the convective blueshift of the \ion{Fe}{I}\,6173.3\,\AA\ line measured with LARS at ultra-high spectral resolution and accuracy. We focus our study of the convective blueshift on the center-to-limb variation of the line shape and the mean Doppler velocity. 

\paragraph{Center-to-limb variation of the line asymmetry:}
The asymmetry of a spectral line contains important information on atmospheric conditions, the composition of granular and intergranular convection, and the line formation itself. Vertical gradients of Doppler shifts can be extracted by the calculation of the bisector curve of a spectral line profile. The bisector of \ion{Fe}{I}\,6173.3\,\AA\ at the solar disk center was shown in Fig.\,\ref{fig_bisectors_6173_mu10}. It features a conspicuous C-shape with a differential blueshift reaching its maximum of $\mathrm{-320\,m\,s^{-1}}$ in the upper half of the line followed by a decrease to $\mathrm{-180\,m\,s^{-1}}$ at the line minimum. 

\begin{figure}[htpb]
\includegraphics[trim= 5mm 0mm 6mm 0mm, clip, width=\columnwidth]{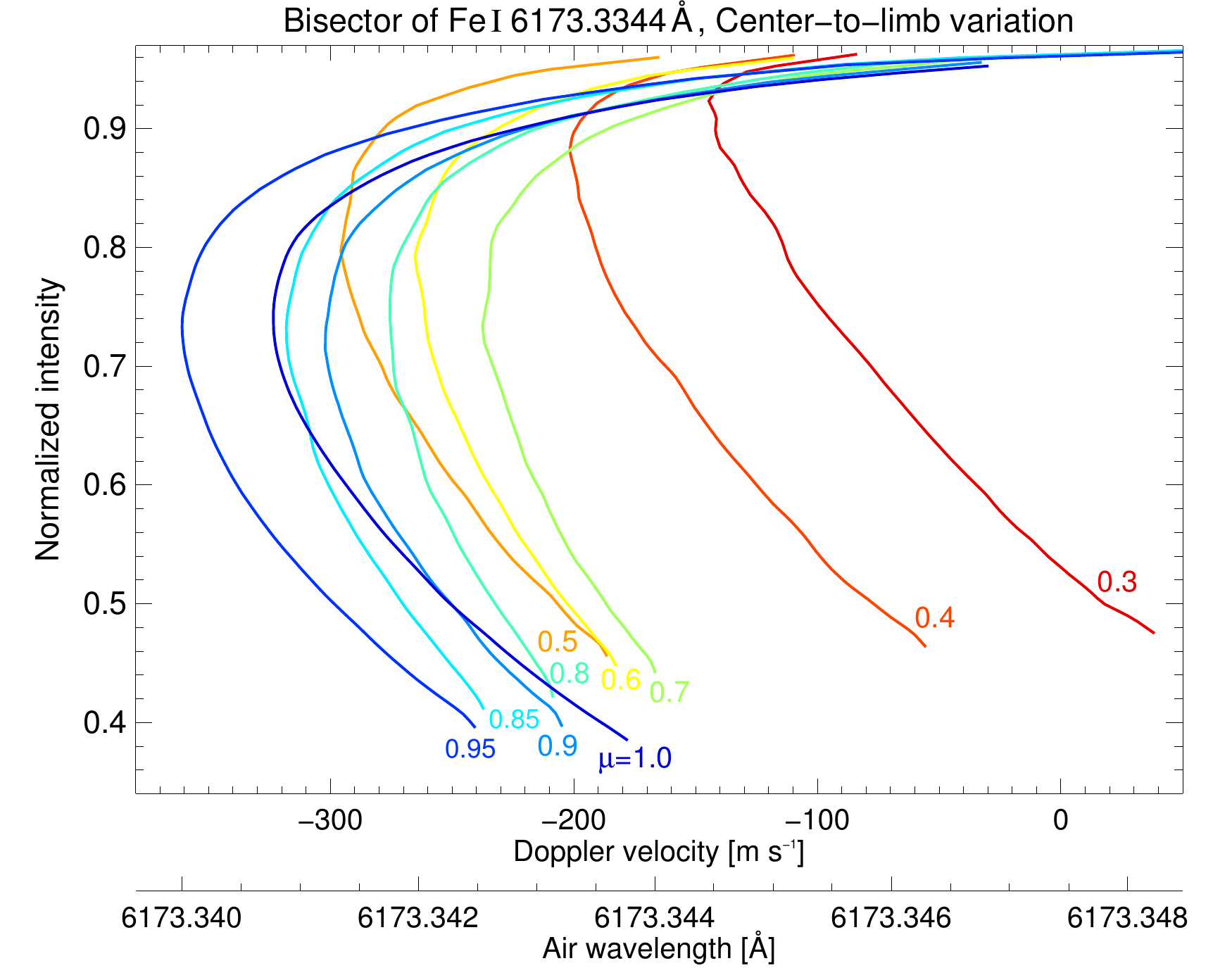}
\caption{Center-to-limb variation of the bisector profile of \ion{Fe}{I}\,6173.3\,\AA\ from $\mu=1.0$ (blue curve) toward $\mu=0.3$ (red curve).  The normalized intensity is displayed against the absolute air wavelength and the Doppler velocity. Each curve represents the average bisector for all measurements at the respective heliocentric position.}
\label{img: shape's comparison}
\end{figure}

In the following, we performed the same bisector analysis for all observed heliocentric positions. We find that the bisector profile of \ion{Fe}{I}\,6173.3\,\AA\ changes significantly from the disk center toward the solar limb. To make a qualitative statement on the global evolution of the line bisector, we verified that the aforementioned scatter caused by p-modes and supergranulation does not affect the systematic shape of the bisector at the respective heliocentric position. We were able to verify that all bisectors at the respective heliocentric position, independently of the observed radial axis, resembled each other in their shape. Differences at each $\mu$-value were only found in the velocity offset. These line shifts reached a scatter up to a few hundred $\mathrm{m\,s^{-1}}$, but did not affect the line asymmetry. Overall, we neither found significant systematic differences between the four radial axis, nor systematic cyclic long-term variations of the convective blueshift. Although, we note that the convective blueshift measured at $\mu=0.8$ and $\mu=0.7$ along the north axis yield predominantly weaker blueshifts. Assuming the velocity offset to be normally distributed, we averaged the bisectors of the four radial axis for each heliocentric position. The systematic evolution of the line asymmetry and Doppler shift of \ion{Fe}{I}\,6173.3\,\AA\ from disk center toward the limb is plotted in Fig.\,\ref{img: shape's comparison}.

In comparison with Paper I (and Fig.\,8 therein), the bisector of the \ion{Fe}{I}\,6173.3\,\AA\ line describes the same shape and center-to-limb variation as the \ion{Fe}{I}\,6302.5\,\AA\ bisector. Both lines were grouped as class 1 bisectors by \citet{1984SoPh...93..219B}. The heliocentric evolution shown in Fig.\,\ref{img: shape's comparison} indicates how fast the bisector shape of \ion{Fe}{I}\,6173.3\,\AA\ changes with increasing distance to the disk center (or decreasing $\mu$-value). From the disk center at $\mu=1.0$ toward $\mu=0.3$ close to the solar limb, the bisector transforms from a distinct C-shape into a ``\textbackslash''-shape. At $\mu=0.95$, the maximum blueshift of $\mathrm{-360\,m\,s^{-1}}$ is reached at a normalized intensity of 0.74. Toward $\mu=0.3$, the maximum blueshift of each bisector profile shifts to $\mathrm{-150\,m\,s^{-1}}$ at an intensity of 0.92. From $\mu=1.0$ to $\mu=0.3$, the line depth decreases by more than 15\%. The normalized intensity of the line minimum increases monotonically from 0.38 at the disk center to 0.48 at $\mu=0.3$. In comparison to the line core velocity of $\mathrm{-180\,m\,s^{-1}}$ at $\mu=1.0$, the blueshift of the line minimum features stronger blueshifts between $\mu=0.95$ and $\mu=0.8$. From $\mu=0.5$ to $\mu=0.3$, the blueshift of the line minimum decreases and even turns into a slight redshift of $\mathrm{+40\,m\,s^{-1}}$.

The C-shaped reversal of the bisector of the spatially averaged line profile is caused by the composition of the {blueshifted line profiles from the bright granules and the redshifted line profiles from the dark intergranular lanes} \citep{1982ARA&A..20...61D}. {The exact interpretation of the physical effects and atmospheric structure causing the line asymmetry would require complex three-dimensional simulations.} At the solar limb, we observe the vertically decreasing blueshift for a larger line-of-sight angle. In addition, horizontal granular flows at higher photospheric layers become important. At  large heliocentric angles and higher photospheric altitudes, the apparently larger fraction of horizontal granular flows away from the observer can cause the observed redshifts at the solar limb \citep{1985SoPh...99...31B}.

\paragraph{Quantitative analysis of the center-to-limb variation of the convective blueshift:}
To perform a more quantitative study of the convective blueshift, we calculated the average Doppler shift of a defined segment of the bisector profile. Due to the asymmetry of the bisectors, we computed the average for three different line segments: the lower 5\,\%, the lower 25\,\%, and the {lower 80\,\% of the line bisector.}

\begin{figure}[htpb]
\includegraphics[trim= 3mm 1mm 5mm 0mm, clip, width=\columnwidth]{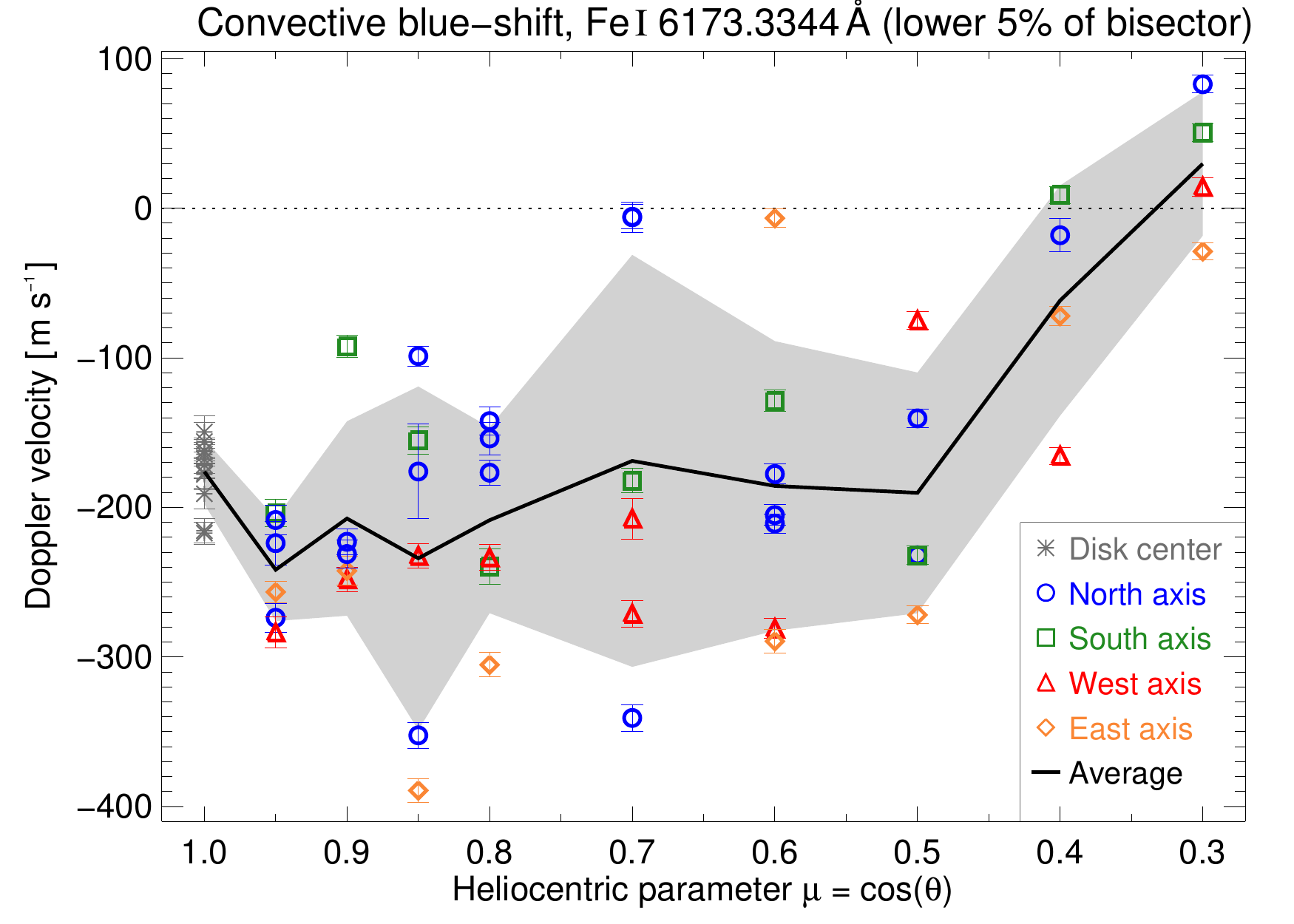}
\caption{Center-to-limb variation of the convective blueshift of \ion{Fe}{I}\,6173.3\,\AA. The Doppler velocities represent the average value of the lower 5\,\% of the bisector profile. Each entry represents the temporal average of an observation sequence, with error bars indicating the mean error. The measurements along the four radial axes can be distinguished by the colors and symbols. The average center-to-limb variation is displayed as black solid line, with its standard deviations as gray area.}
\label{COG Plot}
\end{figure}

The distribution of Doppler velocities for the lower 5\,\% of the bisector is shown in Fig.\,\ref{COG Plot}. Across the solar disk, velocities range from blueshifts of $\mathrm{-390\,m\,s^{-1}}$ to redshifts of $\mathrm{+80\,m\,s^{-1}}$. The mean error of each measurement was of the order of $\mathrm{10\,m\,s^{-1}}$. The standard deviation (indicated by the gray area) around the average Doppler shift highlights the scatter at each heliocentric position. Besides, a systematic trend could be identified for the center-to-limb variation. We calculated the average velocity at each heliocentric position and plotted the mean curve. The convective blueshift of $\mathrm{-180\,m\,s^{-1}}$ at disk center features a slight increase to $\mathrm{-230\,m\,s^{-1}}$ at $\mu=0.85$, with a local minimum at $\mu=0.95$. From $\mu=0.85$ toward $\mu=0.5$, the blueshift decreases again to the velocity measured at disk center. When approaching the solar limb, the blueshift starts to decrease rapidly and even turns into a slight redshift of $\mathrm{+30\,m\,s^{-1}}$ at $\mu=0.3$. This trend of decreasing blueshifts toward the limb confirms the theoretical picture \citep{1985SoPh...99...31B}. There is only a slight increase in blueshift from disk center toward a heliocentric position around $\mu=0.8$ as predicted by theoretical syntheses \citep[e.g.,][]{2011A&A...528A.113D}. However, we must note that the distribution of velocities exhibits the largest scatter for heliocentric positions between  $\mu=0.85$ and $\mu=0.6$. At these locations, solar oscillations and horizontal granular or supergranular flows introduce the largest total error from solar activity on the measurement of the convective blueshift. At disk center, the temporal average reduced the error caused by acoustic oscillations. The scatter around $\mathrm{-180\,m\,s^{-1}}$ is small since horizontal flows have no impact, given by the orthogonal line of sight. Toward the solar limb, the effect of largely vertical p-mode oscillation is therefore small. We conclude from the relatively narrow velocity distribution at the limb position ($\mu=0.5-0.3$) that the spatial averaging successfully minimized the effect of supergranular flow fields.

We compared the velocity distribution based on the analysis of the lower 5\,\% with the results obtained for the lower 25\,\% or 80\,\% of the spectral line. The latter results were shown in Fig.\,\ref{img: HMI comparison}. At first sight, the distribution and mean evolution of the convective blueshift seem much alike for the three different segments. However, there are two major differences. Including larger fractions of the bisector profile leads to an overall increase of the convective blueshift for all measurements. This increase amounts to around $\mathrm{30\,m\,s^{-1}}$ from 5\,\% to 25\%, and additional $\mathrm{40\,m\,s^{-1}}$ from 25\,\% to 80\%. By increasing the averaged bisector segment, we include the stronger blueshifts of the deeper photosphere (at higher intensities as displayed in Fig.\,\ref{img: shape's comparison}). Thus, this trend was to be expected. The second and more subtle change manifests in the gradient of the center-to-limb curve from $\mu=1.0$ toward $\mu=0.5$. As described above and displayed in  Fig.\,\ref{COG Plot}, the center-to-limb variation computed for the lower 5\,\% segment exhibits a slight increase in convective blueshift toward heliocentric positions around $\mu=0.85$. Including a larger portion of the bisector by the 25\,\% or 80\,\% segment changes the gradient of the center-to-limb variation to a more steady decrease in blueshift with decreasing $\mu$-value.

\section{Conclusions}\label{sec_conclusions}
The high spectral resolution and absolute wavelength calibration of LARS enabled measurements of the line shape and Doppler shift of the spectral lines in the 6173\,\AA\ region with an unprecedented accuracy. We investigated the convective blueshift and its center-to-limb variation. A detailed bisector analysis revealed the line shift and asymmetry of the \ion{Fe}{I}\,6173.3\,\AA\ line with an uncertainty of around $\mathrm{5\,m\,s^{-1}}$. 

With a bisector curve of \ion{Fe}{I}\,6173.3\,\AA\ transforming from a C-shape at disk center to a \textbackslash-shape toward the solar limb, we refined former measurements of line asymmetries \citep{1984SoPh...93..219B,1985SoPh...99...31B} and confirmed the change of the convective signature in the shape of the photospheric spectral line. The line core of \ion{Fe}{I}\,6173.3\,\AA\ featured a blueshift of around $\mathrm{-180\,m\,s^{-1}}$. From disk center toward a heliocentric position of $\mu=0.5$, the convective blueshift remained almost stable with a slight increase toward $\mu=0.85$. Toward the solar limb, the blueshift decreased rapidly and even turned into a slight redshift of $\mathrm{+30\,m\,s^{-1}}$. 

Computing the  average Doppler shift of the lower 80\% of the line bisector allowed us a comparison with HMI Doppler velocities obtained with the same spectral line. We find that the slope of the relative center-to-limb variation measured with HMI differs significantly from the absolute values observed with LARS up to $\mathrm{255\,m\,s^{-1}}$. The difference became more apparent for smaller $\mu$-values. By performing a convolution on the LARS data down to the resolution of HMI and applying the HMI method to derive Doppler velocities we showed that the HMI method is not the cause of the discrepancy. The slight differences resulting of different observation and evaluation methods do not explain the significant deviation of the centre-to limb variation of HMI compared to LARS. We conclude that HMI Dopplergrams contain an additional center-to-limb shift which may be caused by intrinsic instrumental effects of HMI, like interference fringes. A potential reduction of those interference patterns will increase the precision of the synoptic HMI Dopplergrams. Supposing that the recalibrated Doppler velocities reproduce the slope of the actual center-to-limb variation of the convective blueshift, we argue that HMI Dopplergrams could be calibrated to an absolute velocity scale with the accurate convective blueshifts measured by LARS taken as reference. With an Dopplergram accuracy of a few ten $\mathrm{m\,s^{-1}}$, we could unambiguously identify weak up- and down-flows in active and quiet Sun regions close to disk center. Moreover, we would be able to directly detect the large-scale meridional flow at the solar surface. Thus, we aim to increase the significance of the LARS reference values by performing more observations with LARS at the VTT. We do not expect any significant change of the average center-to-limb variation, but presume to smoothen the slope curve especially in the range between $\mu=0.95$ and $0.5$.

\begin{acknowledgements} We thank our colleagues at the Kiepenheuer Institute for Solar Physics, at Menlo Systems GmbH, and at the Max Planck Institute of Quantum Optics for their work on the LARS instrument. We especially acknowledge Hans-Peter Doerr for his work on the LARS prototype and his support on the operation of the instrument. We thank Rolf Schlichenmaier and Nazaret Bello Gonz\'alez as co-investigator, and Oliver Wiloth and Michael Wei{\ss}sch\"adel for their assistance during the observation campaigns. The operation of the Vacuum Tower Telescope at the Observatorio del Teide on Tenerife was performed by the Kiepenheuer Institute for Solar Physics Freiburg, which is a public law foundation of the State of Baden-W\"urttemberg. HMI data were provided by the Joint Science Operation Centre (JSOC) and used by courtesy of the NASA/SDO and HMI science teams. This work was funded by the Deutsche Forschungsgemeinschaft (DFG, Ref.-No. Schm-1168/10). We thank Rolf Schlichenmaier for his fruitful comments on the manuscript.
\end{acknowledgements}

\bibliographystyle{aa} 
\bibliography{LARS} 

\begin{thebibliography}{21}
\expandafter\ifx\csname natexlab\endcsname\relax\def\natexlab#1{#1}\fi

\bibitem[{{Acton}(1996)}]{Acton1996}
{Acton}, C.~H. 1996, Planetary and Space Science, 44, 65

\bibitem[{{Asplund}(2005)}]{2005ARA&A..43..481A}
{Asplund}, M. 2005, \araa, 43, 481

\bibitem[{{Balthasar}(1984)}]{1984SoPh...93..219B}
{Balthasar}, H. 1984, \solphys, 93, 219

\bibitem[{{Balthasar}(1985)}]{1985SoPh...99...31B}
{Balthasar}, H. 1985, \solphys, 99, 31

\bibitem[{{Bruls} {et~al.}(1991){Bruls}, {Lites}, \&
  {Murphy}}]{1991sopo.work.....N}
{Bruls}, J.~H.~M.~J., {Lites}, B.~W., \& {Murphy}, G.~A. 1991, in Solar
  Polarimetry, ed. L.~J. {November}, 444

\bibitem[{{Cabrera Solana} {et~al.}(2005){Cabrera Solana}, {Bellot Rubio}, \&
  {del Toro Iniesta}}]{2005A&A...439..687C}
{Cabrera Solana}, D., {Bellot Rubio}, L.~R., \& {del Toro Iniesta}, J.~C. 2005,
  \aap, 439, 687

\bibitem[{{Couvidat} {et~al.}(2012{\natexlab{a}}){Couvidat}, {Rajaguru},
  {Wachter}, {Sankarasubramanian}, {Schou}, \&
  {Scherrer}}]{2012SoPh..278..217C}
{Couvidat}, S., {Rajaguru}, S.~P., {Wachter}, R., {et~al.} 2012{\natexlab{a}},
  \solphys, 278, 217

\bibitem[{{Couvidat} {et~al.}(2012{\natexlab{b}}){Couvidat}, {Schou}, {Shine},
  {Bush}, {Miles}, {Scherrer}, \& {Rairden}}]{2012SoPh..275..285C}
{Couvidat}, S., {Schou}, J., {Shine}, R.~A., {et~al.} 2012{\natexlab{b}},
  \solphys, 275, 285

\bibitem[{{de la Cruz Rodr{\'{\i}}guez} {et~al.}(2011){de la Cruz
  Rodr{\'{\i}}guez}, {Kiselman}, \& {Carlsson}}]{2011A&A...528A.113D}
{de la Cruz Rodr{\'{\i}}guez}, J., {Kiselman}, D., \& {Carlsson}, M. 2011,
  \aap, 528, A113

\bibitem[{{Doerr}(2015)}]{Doerr2015}
{Doerr}, H.-P. 2015, PhD thesis, University of Freiburg

\bibitem[{{Dravins}(1982)}]{1982ARA&A..20...61D}
{Dravins}, D. 1982, \araa, 20, 61

\bibitem[{Kramida {et~al.}(2015)Kramida, {Yu.~Ralchenko}, Reader, \& {and NIST
  ASD Team}}]{NIST_ASD}
Kramida, A., {Yu.~Ralchenko}, Reader, J., \& {and NIST ASD Team}. 2015, {NIST
  Atomic Spectra Database (ver. 5.3), [Online]. Available:
  {\tt{http://physics.nist.gov/asd}} [2017, March 24]. National Institute of
  Standards and Technology, Gaithersburg, MD.}

\bibitem[{{L{\"o}hner-B{\"o}ttcher} \&
  {Schlichenmaier}(2013)}]{Loehner+Schlichenmaier2013}
{L{\"o}hner-B{\"o}ttcher}, J. \& {Schlichenmaier}, R. 2013, \aap, 551, A105

\bibitem[{{L{\"o}hner-B{\"o}ttcher} {et~al.}(2017){L{\"o}hner-B{\"o}ttcher},
  {Schmidt}, {Doerr}, {Kentischer}, {Steinmetz}, {Probst}, \&
  {Holzwarth}}]{2017A&A...607A..12L}
{L{\"o}hner-B{\"o}ttcher}, J., {Schmidt}, W., {Doerr}, H.-P., {et~al.} 2017,
  \aap, 607, A12

\bibitem[{{L{\"o}hner-B{\"o}ttcher} {et~al.}(2018){L{\"o}hner-B{\"o}ttcher},
  {Schmidt}, {Stief}, {Steinmetz}, \& {Holzwarth}}]{2018A&A...611A...4L}
{L{\"o}hner-B{\"o}ttcher}, J., {Schmidt}, W., {Stief}, F., {Steinmetz}, T., \&
  {Holzwarth}, R. 2018, \aap, 611, A4

\bibitem[{{Nave} {et~al.}(1994){Nave}, {Johansson}, {Learner}, {Thorne}, \&
  {Brault}}]{1994ApJS...94..221N}
{Nave}, G., {Johansson}, S., {Learner}, R.~C.~M., {Thorne}, A.~P., \& {Brault},
  J.~W. 1994, \apjs, 94, 221

\bibitem[{{Neckel}(1999)}]{1999SoPh..184..421N}
{Neckel}, H. 1999, \solphys, 184, 421

\bibitem[{{Schou} {et~al.}(2012){Schou}, {Scherrer}, {Bush}, {Wachter},
  {Couvidat}, {Rabello-Soares}, {Bogart}, {Hoeksema}, {Liu}, {Duvall}, {Akin},
  {Allard}, {Miles}, {Rairden}, {Shine}, {Tarbell}, {Title}, {Wolfson},
  {Elmore}, {Norton}, \& {Tomczyk}}]{2012SoPh..275..229S}
{Schou}, J., {Scherrer}, P.~H., {Bush}, R.~I., {et~al.} 2012, \solphys, 275,
  229

\bibitem[{{Schuck} {et~al.}(2016){Schuck}, {Antiochos}, {Leka}, \&
  {Barnes}}]{2016ApJ...823..101S}
{Schuck}, P.~W., {Antiochos}, S.~K., {Leka}, K.~D., \& {Barnes}, G. 2016,
  Astrophysical Journal, 823, 101

\bibitem[{{Snodgrass} \& {Ulrich}(1990)}]{1990ApJ...351..309S}
{Snodgrass}, H.~B. \& {Ulrich}, R.~K. 1990, \apj, 351, 309

\bibitem[{{Stix}(2002)}]{2002tsai.book.....S}
{Stix}, M. 2002, {The sun: an introduction}

\end{thebibliography}


\end{document}